\documentclass[pra,nofootinbib,a4paper,onecolumn,notitlepage]{revtex4-1}
\usepackage[english]{babel}

\pdfpagewidth=\paperwidth
\pdfpageheight=\paperheight

\usepackage[latin9]{inputenc}
\usepackage{textcomp}
\usepackage{amstext}

\usepackage{amsmath}
\usepackage{amsbsy}
\usepackage{amsthm}
\usepackage{amstext}
\usepackage{amsxtra}
\usepackage{amscd}
\usepackage{amsopn}
\usepackage[english]{babel}
\usepackage{babel}
\usepackage{mathtools}
\usepackage{epstopdf}
\usepackage{psfrag,graphicx}
\usepackage{graphicx}
\usepackage{epsf}
\usepackage{epstopdf}
\usepackage{color}
\usepackage{dsfont}
\usepackage{upgreek}
\usepackage{tipa}
\usepackage{phonetic}

\begin{document}

\title{Device-Independent Bit Commitment based on the CHSH Inequality}

\author{N. Aharon\textsuperscript{1,2}, S. Massar\textsuperscript{3}, S. Pironio\textsuperscript{3},
and J. Silman\textsuperscript{3}}

\affiliation{\textsuperscript{1}School of Physics and Astronomy, Tel-Aviv University,
Tel-Aviv 69978, Israel \\
\textsuperscript{2}{Racah Institute of Physics, The Hebrew University of Jerusalem,
Jerusalem 91904, Israel} \\
 \textsuperscript{3}Laboratoire d'Information Quantique, Universit\'{e}
Libre de Bruxelles (ULB), 1050 Bruxelles, Belgium}

\begin{abstract}
\textbf{Bit commitment and coin flipping occupy a unique place in the device-independent landscape, as the only device-independent protocols thus far suggested for these tasks are reliant on tripartite GHZ correlations. Indeed, we know of no other bipartite tasks, which admit a device-independent formulation, but which are not known to be implementable using only bipartite nonlocality.  Another interesting feature of these protocols is that the pseudo-telepathic nature of GHZ correlations -- in contrast to the generally statistical character of nonlocal correlations, such as those arising in the violation of the CHSH inequality -- is essential to their formulation and analysis. In this work, we present a device-independent bit commitment protocol based on CHSH testing, which achieves the same security as the optimal GHZ-based protocol. The protocol is analyzed in the most general settings, where the devices are used repeatedly and may have long-term quantum memory. We also recast the protocol in a post-quantum setting where both honest and dishonest parties are restricted only by the impossibility of signaling, and find that overall the supra-quantum structure allows for greater security.}
\end{abstract}

\maketitle

\section{Introduction}

The security of cryptographic protocols, whether quantum or classical,
depends on the satisfaction of certain assumptions. These include
the integrity of each party's lab and their having a trusted source
of randomness to make the random choices called for by the protocol.
Beyond these, \textit{classical} protocols will also usually include
assumptions regarding the computational power of dishonest parties.
The security of \textit{quantum} protocols, in contrast, is based
only on the validity of quantum theory. Nevertheless, to harness this
validity, assumptions regarding the implementation must be made. Most protocols make many such assumptions, including regarding the internal workings of the devices used in the implementation, e.g. specifying the Hilbert space dimension of
the quantum systems used and the bases of the measurements performed. Protocols of this type are said to be \textit{device-dependent}. Clearly,
it is desirable to base security on a minimum number of assumptions,
as this facilitates its evaluation. The aim of the \textit{device-independent}
 approach to quantum cryptography \cite{Mayers04,Barrett05} is
to do just that by doing away with a maximum number of assumptions
regarding the implementation.

More specifically, a cryptographic protocol is said to be device-independent if its
security can be guaranteed without making assumptions about the
internal workings of the devices used in its implementation. This can be achieved by carrying out Bell tests on entangled systems. The level
of security is then deduced from the observed amount of nonlocality. In particular,
each device is treated as a black box with knobs and registers for
selecting and displaying (classical) inputs and outputs.
For instance, in device-independent quantum key-distribution a high violation of the CHSH
inequality \cite{Clauser69} guarantees that an eavesdropper will have no information
about the (post-processed) key \cite{Barrett05,Acin07,Pironio09,McKague09,Masanes11,Reichardt13A,Pironio13B,Vazirani12B}.
In contrast, in the (device-dependent) entanglement-based version of the
BB84 protocol it has been shown that if the source dispenses qudits instead of qubits
then security can be utterly compromised \cite{Magniez06,Acin06}. Indeed, recent hacking attacks
on quantum key-distribution systems, such as those of \cite{Xu10,Lydersen10},
exploit device-dependent modes of failure and would not be successful
in device-independent settings.\\

In addition to quantum key-distribution,
device-independent protocols have been introduced for diverse tasks such as randomness generation \cite{Colbeck07,Pironio10A,Pironio13A,Fehr13,Vazirani12A,MillerShi,Coudron}, 
the self-testing of quantum computers \cite{Mayers04,Magniez06,McKague11,Reichardt13A}, state estimation \cite{Bardyn09,McKague12,Yang12,Yang13,Bamps},
genuine multipartite entanglement certification \cite{Bancal11}, and entanglement quantification \cite{Moroder13}. However, until recently it was
not known whether the scope of the device-independent approach also covers protocols
in the \textit{distrustful cryptography class}, where the parties
do not trust each other and may have conflicting goals. Problems in this class
present us with an extra challenge in device-independent settings
as compared to tasks such as quantum-key distribution. Namely, how to allow remote
distrustful parties to certify the presence of nonlocality without collaborating. In \cite{Silman11} it was shown
that imperfect bit commitment\footnote{While quantum mechanics does not allow for perfect bit commitment \cite{Lo97,Mayers97}, imperfect bit commitment is nevertheless possible \cite{Spekkens01,Chailloux11}.}
 admits a device-independent formulation, and, since bit commitment may serve as a primitive for coin flipping, so does coin flipping (a device-independent coin flipping protocol, not based on bit commitment, was also introduced in \cite{Aharon14}). Whether these results extend to all problems in the distrustful cryptography class remains an open question.

A notable feature of the protocols of \cite{Silman11,Aharon14} is that they are based on GHZ correlations \cite{Greenberger89,Mermin90}. Indeed, bit commitment and coin flipping are the only examples we have of bipartite tasks, which admit a device-independent formulation, but which are not known to admit one based on CHSH testing (i.e. sequential tests of the CHSH inequality), or, more generally, on some other bipartite Bell inequality testing. This is especially interesting in light of Reichardt et al.'s  recent demonstration \cite{Reichardt13A}  that CHSH testing can provide the basis for many device-independent applications in the most general settings where the devices have long-term quantum
memory.

In \cite{Silman11,Aharon14}, the pseudo-telepathic nature of GHZ correlations is exploited to circumvent the unique difficulties associated with distrustful cryptography, specifically, the fact that different parties have conflicting goals and do not trust each other. 
Quantum pseudo-telepathy is the term coined for the phenomenon of always winning in nonlocal
games, which classically (i.e. without sharing entanglement) can only be won part of the time. A famous example
is the GHZ game \cite{Vaidman99}. In particular, pseudo-telepathy
entails perfect correlations. In \cite{Silman11} pseudo-telepathy
is used to allow Bob to verify the presence of nonlocal correlations
(GHZ correlations) and \textit{at the same time} to verify Alice's
commitment (that the token of her commitment is consistent with the
value of the bit she reveals). Crucially, Bob uses the same measurements
to verify both the presence of nonlocality and the commitment.

Unfortunately, pseudo-telepathy is absent in the CHSH setting \cite{Gisin07}, and so it is a priori unclear whether bipartite distrustful cryptographic tasks can be based on CHSH testing -- which is the case for all other examples of bipartite tasks that are known to admit a device-independent formulation. Beyond a theoretical interest, this question is also practically motivated, since manipulating tripartite entanglement, as would be required in a GHZ-based protocol, is obviously more difficult than manipulating EPR pairs, as would be required in a CHSH-based protocol.\\

In this work we present a device-independent bit commitment protocol, based on CHSH
testing, which achieves the same security as that of \cite{Silman11}: (In the limit of an infinite number of tests) Alice's control equals $\cos^2(\frac{\pi}{8})\simeq 0.8536$, while Bob's information gain equals $0.75$. 
This shows that pseudo-telepathy is not only inessential for device-independent distrustful
cryptography, but that its absence does not necessarily impact security. Specifically, we show how to guarantee that the devices have no way of telling whether they are used as part of the nonlocality
testing phase or the verification of the commitment; this being the crucial element
on which security hinges. 

Our security analysis covers the case of imperfect devices (i.e. the CHSH inequality is not maximally violated) and is carried out in the most general settings where memory effects (the dependence
of a measurement outcome not only on the setting, but also on previous
settings and outcomes) are taken into account.

It should be noted that in our protocol the reveal time is fixed and cannot be chosen at will by Alice. Strictly speaking, the protocol is thus not a bit commitment protocol.  Nevertheless, depending on the application, it may still be used as a primitive. For example, our protocol can be used to implement coin flipping. The restriction on the reveal time can be lifted at the price of increasing Alice's cheating probability (see Appendix B), or by working in the large office scenario where instead of a pair of boxes there are many pairs (see Appendix C).  

We also study the problem in a post-quantum world where \textit{both
dishonest and honest} parties are restricted only by the impossibility of signaling.
This helps us identify the contribution of different resources
to security. On the one hand, we might expect such a world to offer
less security since a dishonest party would have access to stronger
correlations. On the other hand, we might expect the converse, since
the protocol itself could be modified to make use of these stronger
correlations (in particular, pseudo-telepathy is restored in this
setting). It turns out that on the balance this allows for more security.\\

The paper is structured as follows. We begin in Section II by defining
the problem of bit commitment, and making explicit exactly what we mean
by device-independence. Next, in Section III, we present the
protocol, followed by the proofs of Alice's and Bob's securities in
Sections IV and V, respectively. We conclude with a summary in Section VI. In  Appendix A we present the post-quantum version of our protocol. Appendices B and C present modifications of the protocol where Alice can freely choose the reveal time.

\section{Background}

\subsection{Bit commitment}

Bit commitment is a cryptographic primitive comprising two remote, distrustful parties. Party $\mathcal{A}$, usually referred to as Alice, commits a bit to party $\mathcal{B}$, usually referred to as Bob, such that following her commitment Alice cannot change its value and Bob is unable to learn it until she chooses to reveal it. Classically, if the dishonest party's computational power is unlimited, they can cheat perfectly. Quantumly, the dishonest party cannot cheat perfectly \cite{Spekkens01}, though perfect bit commitment is still impossible \cite{Lo97,Mayers97}.

A bit commitment protocol consists of two phases: the commit phase in which Alice sends Bob some token of her commitment, and the reveal phase in which Alice reveals to Bob the value of the committed bit. The probability with which dishonest Alice is able to control the value of the bit she wants to reveal following the commit phase, without being caught cheating by Bob, is referred to as Alice's control, which we will denote by $P_\mathrm{cont}=\frac{1}{2}(p_0+p_1)$. Here $p_0$ ($p_1$) is Alice's probability of successfully revealing $0$ ($1$) and the factor of $\frac{1}{2}$ is due to the implicit assumption that she is equally likely to wish to reveal $0$ as $1$. Similarly, dishonest Bob's probability of correctly learning the value of the bit before the reveal phase is referred to as Bob's information gain, which we will denote by $P_\mathrm{gain}$. In a perfect bit commitment protocol $P_\mathrm{cont}=P_\mathrm{gain}=\frac{1}{2}$. A protocol is said to be balanced if $P_\mathrm{cont}=P_\mathrm{gain}$. Quantumly, in any balanced protocol $P_\mathrm{cont}=P_\mathrm{gain} \gtrsim 0.739$, with the bound being saturable \cite{Chailloux11}.

\subsection{Device-independence}

In this subsection we make more concrete exactly what we mean by device-independence. We make the following standard assumptions.
\begin{enumerate}
\item Alice and Bob have access to boxes, each with
a knob for selecting a classical input $s$ and a register for displaying a classical output $r$.
Entering an input always results in an output (i.e.
we do not consider losses).
\item Alice and Bob, whether honest or dishonest, are restricted by quantum
theory.
\item The boxes may be prevented at will from communicating with one another.
\item Alice and Bob each have a trusted source of randomness.
\item No information leaks out of an honest party's lab.
\end{enumerate}

Suppose now that an honest party has a pair of boxes $0$ and $1$. Assumptions 2 and 3 imply that the
 probability of
outputting $r^0$ and $r^1$  when inputting $s^0$ and $s^1$ into boxes $0$ and $1$, respectively, is given by
\begin{equation}
P\left(r^0,\,r^1 | s^0,\,s^1\right)=\mathrm{Tr}\left(\rho\, \Pi_{r^0|s^0} \otimes \Pi_{r^1|s^1}\right)\,,
\end{equation}
where $\rho$ is some joint quantum state and $\Pi_{r^i|s^i}$ is the POVM element corresponding to inputting $s^i$ and
outputting $r^i$. This is the \emph{only} constraint on the boxes' behavior. Specifically, a dishonest party may choose the state $\rho$
and the POVM elements $\Pi_{r^i |  s^i}$ as best suits them. The boxes may also have internal memories, clocks, gyroscopes,
etc., allowing a dishonest party to program
them such that their behavior depends on their location, their past trajectories, the time at which inputs are fed, or any
other aspect of their past history.

In the following, we will consider situations where boxes are sent from one party to the other. By this, we do not mean that actual measurement devices are
sent (though it is easier to present and formulate our results in this way). Instead, what we mean is that quantum states, or classical information,
encoding instructions for the measurement devices, are exchanged between the parties, such that \emph{in an honest execution of the protocol} the same state $\rho$ and the POVM elements
$\Pi_{r^i|s^i}$ characterizing the behavior, say, of Alice's box before the transmission of quantum information, will characterize the
behavior of Bob's box after receiving the transmission.

Finally, we wish to emphasize that spacelike related measurements are not necessary in order to prevent the boxes from communicating (i.e. assumption 3). We may equally well shield each box (see  \cite{Pironio09,Pironio10A} for a discussion of this point). This observation is important because $(i)$ in our
protocol many of the measurements are not spacelike related; $(ii)$ relativistic causality is by itself sufficient for perfect bit commitment (whether purely classical \cite{Kent99} or quantum \cite{Kent15}), albeit at the cost
of assigning at least one party two remote secure labs.

\section{The Protocol}

Before we go on to present the protocol, we fix notation.
In the following we will consider a pair of four-input, $\{0,\,1,\,2,\,3\}$, two-output, $\{0,\,1\}$, boxes. The random variables designating the input and output corresponding to the $k\,$th use of box $i$ will be labeled by $S^i_k$ and $R^i_k$, respectively, with a specific realization (i.e. a specific value which they may assume) being labeled by lower-case letters $s^i_k$ and $r^i_k$. Similarly, the random strings corresponding to $k$ consecutive uses of box $i$ will be labeled by $\mathrm{\mathbf{S}}^i_k$ and $\mathrm{\mathbf{R}}^i_k$. We define  $W_k=\{S^0_k,\,S^1_k,\,R^0_k, \,R^1_k\}$ and $\mathbf{W}_k=\{W_1,\,\dots,\,W_k\}$. We will refer to a specific realization $\mathbf{w}_k=\{w_1,\,\dots,\,w_k\}$ as the history of the protocol. Finally, $|0 \rangle$ ($|1 \rangle$) will be taken to represent the positive (negative) eigenstate of $\sigma_z$.

The protocol is based on EPR-state correlations. In an ideal implementation, the boxes are supposed to give rise to a violation of $2\sqrt{2}$ of the CHSH inequality in the sense that
\begin{equation}
\sum_{r^0_n,\,r^1_n,\,s^0_n,\,s^1_n=0,\,1}(-1)^{r^0_n\oplus r^1_n \oplus s^0_n s^1_n} P(r^0_n,\,r^1_n|s^0_n,\,s^1_n)=2\sqrt{2}\qquad \forall n\,.
\end{equation}
In addition, the boxes are supposed to output $r^0_n=r^1_n$ given the pairs of inputs $s^0_n=i$, $s^1_n=i+2 \mod 4$ ($i=0,\,1,\,2,\,3$).
These correlations can be quantumly realized by preparing $N$ qubits, each in the $|\phi^+\rangle=\frac{1}{\sqrt{2}}(|00\rangle+|11\rangle)$ state. The inputs $0$, $1$, $2$, and $3$ of box $0$ correspond to the measurements $\sigma_x$, $\sigma_z$, $\sigma_{\pi/4}$, and $\sigma_{3\pi/4}$, respectively, where $\sigma_\theta=\cos\theta \sigma_z+\sin\theta \sigma_x$. The inputs $0$, $1$, $2$, and $3$ of box $1$ correspond to the  measurements $\sigma_{\pi/4}$, $\sigma_{3\pi/4}$, $\sigma_x$, and $\sigma_z$, respectively.

Since we would also like to consider the noisy case, we do not assume in the following that the parties have perfect resources, i.e. the boxes are expected to give rise to a CHSH violation $I<2\sqrt{2}$ and the outcomes $r^0_n$, $r^1_n$ for the pairs of inputs $s^0_n=i$ and $s^1_n=i+2 \mod 4$ are not perfectly correlated.

We consider a family of protocols. Each protocol in the family is specified by a parameter $N>1$ and a series of fixed times $t_i$ $(i=1,\ldots,N+1)$ with $t_{i-1}<t_i<t_{i+1}$. For a given $N$ and choice of $t_i$, the protocol proceeds as follows:
\begin{enumerate}
\item \emph{Random selection} -- At time $t^a <t_1$ Bob picks uniformly at random, and \emph{in private}\footnote{It is crucial that $n$ is chosen privately and randomly by Bob in
order for him to be able to ascertain that in the $n+1\,$th use the boxes are CHSH violating. Otherwise,
dishonest Alice can prepare the boxes such that they are CHSH violating only in the first $n$ uses.
}, a number $n\in\{1,\,\dots,\,N\}$  and two input strings $\mathrm{\mathbf{s}}^0_n\in\{0,\,1\}^n$ and $\mathrm{\mathbf{s}}^1_n\in\{0,\,1\}^n$. At each of the $n$ times $t_i$ he feeds $s^0_i$ and $s^1_i$ into boxes $0$ and $1$, respectively.
He uses the corresponding output strings $\mathrm{\mathbf{r}}^0_n$ and $\mathrm{\mathbf{r}}^1_n$ to compute the observed CHSH violation
\begin{equation}\label{chsh}
\bar{I}_n\left(\mathrm{\mathbf{w}}_n\right)=\frac{1}{n}\sum_{k=1}^n I\left(  w_k  \right)\,.
\end{equation}
where
\begin{equation}\label{chsh_k}
I(W_k)=4\sum_{r^0_k,\,r^1_k,\,s^0_k,\,s^1_k=0,1}(-1)^{r^0_k\oplus r^1_k \oplus s^0_k s^1_k}\delta_{R^0_k r^0_k}\delta_{R^1_k r^1_k}\delta_{S^0_k s^0_k}\delta_{S^1_k s^1_k}\,,
\end{equation}
is the CHSH indicator function\footnote{The factor of $4$ in (\ref{chsh_k}) is just a normalization taking into account that Bob picks all four possible input pairs with equal probability. With this definition the expected value $E(I(W_k))=\sum_{r^0_k,\,r^1_k,\,s^0_k,\,s^1_k=0,1}(-1)^{r^0_k\oplus r^1_k \oplus s^0_k s^1_k}P(r^0_k\,r^1_k| s^0_k\, s^1_k)$ is the usual CHSH expression.} at step $k$. Bob then compares $\bar{I}_n\left(\mathrm{\mathbf{w}}_n\right)$ to some previously agreed threshold $I_\mathrm{th}$.
If $\bar{I}_n(\mathrm{\mathbf{w}}_n) < I_\mathrm{th}$ he aborts the protocol.
Otherwise, he flips a classical coin. Denote its outcome by $c$. At time $t^b  < t_{n+1}$ he
sends box $c$ to Alice.
\item \emph{Commit phase} -- Let $b$ be the value of the bit Alice wishes to commit. Alice inputs $s^c_{n+1}=b+2$ into her
box. She selects uniformly at random a classical bit $a$, and at time $t^c$ ($t^b<t^c< t_{n+1}$)
sends Bob the classical bit $q=r^c_{n+1}\oplus a b$ as a token
of her commitment.
\item \emph{Reveal phase} -- At time $t^d$ ($t^c<t^d< t_{n+1}$) Alice sends Bob $b$
and $r^c_{n+1}$. If $b$ and $r^c_{n+1}$ are not received before $t_{n+1}$ Bob aborts. Otherwise, Bob checks whether $q=r^c_{n+1}$ or $q=r^c_{n+1}\oplus b$. If both relations are not satisfied, he aborts. Else 
at time $t_{n+1}>t^d $ he inputs $s^{\bar{c}}_{n+1}=s^c_{n+1}-2=b$ into his box and verifies that $r^{\bar{c}}_{n+1}=r^c_{n+1}$.
If this last test fails, he aborts.
\end{enumerate}

We note that in an honest execution of the protocol, Bob may end up aborting the protocol in the random selection phase \emph{even when the settings are ideal}. Moreover, Bob will never know if his having to abort is due to Alice having been dishonest or due to the boxes having exhibited a statistically unlikely behavior. This is just a by-product of the statistical nature of the protocol, which is of course absent in the limit that $N\rightarrow\infty$. However, if Bob does not abort the protocol in the random selection phase, then (assuming ideal settings and an honest execution) he will not abort it in the reveal phase and will always learn the correct value of the bit committed by Alice. In contrast, if there are physical imperfections present, such as noise or a misalignment of the measurement axes, then there is a non-vanishing probability that Bob will abort the protocol in the reveal phase even when Alice is honest. This is true of any practical formulation (i.e. a formulation accommodating imperfections), and has nothing to do with the protocol being device-independent. 

We have required that Bob's measurements, including the one in the reveal phase, take place at fixed times $t_i$\footnote{In fact, it is sufficient to only require that measurement $i+1$ takes place at any time during the interval $(t_i,\,t_{i+1}]$, so long as this time is large enough to allow for the commit and reveal  phases to be completed in the remainder of the interval.}. This is in order to ensure that the box Bob keeps cannot tell whether it is being measured in the random selection phase or in the reveal phase (unless Bob picked $n=N$). Otherwise, Alice may program the boxes such that in the random selection phase they maximally violate the CHSH inequality, while in the reveal phase they behave deterministically, thereby allowing her to cheat perfectly. Specifically, the intervals $t_{i+1}-t_i$ must be sufficiently long to allow the following sequence of operations: (\emph{i}) the sending of quantum information from Bob to Alice, (\emph{ii}) Alice's measurement of the quantum system received from Bob, (\emph{iii}) the sending of classical information from Alice and its receipt by Bob, and (\emph{iv}) Bob's measurement of the quantum system remaining in his possession at $t_{i+1}$.

As mentioned earlier, since the reveal time cannot be chosen at will, strictly speaking, the protocol is not a bit commitment protocol. Nevertheless, depending on the application, it may still be used as a primitive. For example, our protocol may be used to implement coin flipping. The restriction on the reveal time can be lifted at the price of increasing Alice's control (see Appendix B), or by working in the large office scenario (see Appendix C).  

\section{Alice's security}

In the following, when considering the $n+1\,$th measurement of the boxes, i.e. the measurements taking place in the commit and reveal phases, we drop the subscript $n+1$ on the $s^i_{n+1}$ and $r^i_{n+1}$.

\subsection{Bob's information gain}

Alice only receives a single box from Bob and does not verify
the CHSH violation. Bob's most general cheating strategy  is therefore
to prepare Alice's box in an entangled state with an ancillary system
in his possession. Since in the commit phase Bob receives from Alice a single classical bit $q$, Bob will perform one out of a pair
of two-outcome measurements on his ancillary system to infer
Alice's input $s^c$ (and consequently the committed bit $b=s^c-2$). We denote Bob's binary input and output by $m$ and $g$, where $m=0$ ($m=1$) corresponds
to the measurement he carries out when Alice sends $q=0$ ($q=1$),
and $g$ is his guess of $s^c$. The probability $P\left(g\mid r^c,\, s^c,\,m\right)$
of obtaining the output $g$, given the input
$m$, explicitly depends on Alice's input-output pair $s^c$ and
$r^c$ (or, what is the same thing, on $b$ and $r^c$) because Bob's ancillary system and Alice's box are entangled.
Bob's information gain is therefore given by:

\begin{eqnarray}
P_{\mathrm{gain}} & = & \max_{\mathcal{S}}\sum_{r^c,\, b,\, a=0,\,1}P\left(r^c\mid s^c=b+2\right) P\left(g=b\mid r^c,\, s^c=b+2,\,m=r^c\oplus ab\right)\nonumber \\
 & = & \frac{1}{4}\max_{\mathcal{S}}\!\sum_{r^c,\, b=0,\,1}\! P\left(r^c\mid s^c=b+2\right)\Bigl[P\left(g=b\mid r^c,\, s^c=b+2,\,m=r^c\right)\bigr.\nonumber  \bigl.+P\left(g=b\mid r^c,\, s^c=b+2,\,m=r^c\oplus b\right)\Bigr]\\
 & = & \frac{1}{4}\max_{\mathcal{S}}\sum_{r^c,\, b=0,\,1}\Bigl[P\left(r^c,\, g=b\mid s^c=b+2,\, m=r^c\right)\bigr.\bigl.+P\left(r^c,\, g=b\mid  s^c=b+2,\,m=r^c\oplus b\right)\Bigr]\nonumber \\
 & = & \frac{1}{4}\max_{\mathcal{S}}\Bigl[2P\left(0,\,0\mid 2,\,0\right)+2P\left(1,\,0\mid 2,\,1\right)+P\left(0,\,1\mid 3,\,0\right)\bigr.\bigl.+P\left(1,\,1\mid 3,\,1\right)+P\left(0,\,1\mid 3,\,1\right)+P\left(1,\,1\mid 3,\,0\right)\Bigr]\,,
 \label{Bound Pgain}
\end{eqnarray}
where $\mathcal{S}$ denotes the set of all cheating strategies.
Note that since Alice is honest she picks $b$ and $a$ fully at random and so for any pair $b$ and $a$ $P(b,\,a)=\frac{1}{4}$.
From normalization and the no-signaling constraints (i.e. $\sum_{r^1=0,\,1}P(r^0,\, r^1\mid s^0,\, 0)=\sum_{r^1=0,\,1}P(r^0,\, r^1\mid s^0,\, 1)$
and $\sum_{r^0=0,\,1}P(r^0,\, r^1\mid 2,\, s^1)=\sum_{r^0=0,\,1}P(r^0,\, r^1\mid 3,\, s^1)$) we obtain that $P(s^1,\,0|2,\, s^1)+P(0,\,1|3,\, s^1)+P(1,\,1|3,\, s^1)\leq1$
and $P\left(0,\,0\mid 2,\,0\right)+P\left(1,\,0\mid 2,\,1\right)\leq1$, implying that $P_{\mathrm{gain}} \leq \frac{3}{4}$.

\subsection{Bob's optimal cheating strategy}

Bob's optimal cheating strategy is to prepare Alice's box such that $r^c=s^c-2$ and guess $b=q$. Since Alice is honest $q$ equals $r^c$ (and thus equals $b=s^c-2$) $75\%$ of the time. Alternately, Bob can employ a device-dependent strategy (i.e. where Alice's measurements are those prescribed by the protocol). In this strategy Bob actually prepares the boxes as prescribed by the protocol. Noting that the measurement settings which correspond to $s^{\bar{c}}=0$ and $s^c=2$ are identical, Bob inputs $0$ into his box. Since $q$ equals Alice's
outcome $75\%$ of the time, Bob always treats it as her output. If his outcome equals $q$ Bob guesses that Alice input $2$, otherwise, he guesses that she input $3$. Whenever Alice inputs $2$, Bob's guess is correct. Whenever Alice inputs $3$, Bob's guess is correct only half of the time. Bob's information gain is thus seen to equal the optimum, as well as the result of \cite{Silman11}.

\section{Bob's security}

This section is divided into three. In Subsections A and B we consider the case where the boxes at the  $n+1$th iteration (i.e. after Bob's CHSH estimation) are known to be characterized by a fixed Bell violation $I \geq I_\mathrm{th}$. As we will see in Subsection C, this is equivalent to considering the asymptotic limit in which the number of tests Bob carries out tends to infinity. Specifically, in Subsection A we derive an upper bound on Alice's control, given the CHSH expectation value $I$, and in Subsection B we present an optimal cheating strategy which saturates it.  Finally, in Subsection C we use the bound derived in Subsection A to derive an upper bound on Alice's control in the general case where Bob carries out an arbitrary number of tests. In the limit that this number tends to infinity we recover the bound of Subsection A.

\subsection{Alice's control in the asymptotic limit}

Most generally, in the commit phase Alice carries out a two-outcome measurement on the systems in
her possession: box $c$, which she received from Bob, and possibly some ancillary system with which the boxes may be
entangled. The result of the measurement determines the value of $q$ she sends Bob. In the reveal phase she then performs one out of four possible two-outcome measurements, depending on the value of $q$ and whether she wishes to reveal $0$ or $1$, in order to determine $r^c$.  We note, however, that when she wishes to reveal $0$ the last measurement is redundant because $q$ must equal $r^c$. Alice therefore does not lose anything by \emph{always} performing in the reveal phase one out of the two measurements corresponding to her wishing to reveal $1$. This implies that without loss of generality these two measurements may be combined with the measurement in the commit phase to form a single four-outcome measurement in the commit phase. This measurement decides the two values of $r^c$, and simultaneously the value of $q$. To sum up, in the commit phase Alice carries out a four-element POVM $\mathcal{M}^c=\{M_{kl}^c\}$ ($k,\,l=0,\,1$) acting on $\mathcal{H}^{c}$, such that if she wishes to reveal $0$ ($1$) she sends Bob $q=k$ in the commit phase and $r^c=k$ ($r^c=l$) in the reveal phase.

Suppose that Alice wishes to reveal $b=0$ ($s^c=2$). Bob will first check whether $r^c=q$ (since $b=0$, $r^c\oplus b=r^c$). Bob will then input $s^{\bar{c}}=0$ and verify that $r^{\bar{c}}=r^c$. In this case Alice's cheating probability equals $\frac{1}{2}\sum_{k,\,l=0,\,1}\left[P(r^1=k,\,(k,\,l)|s^1=0,\,\mathcal{M}^0)+P(r^0=k,\,(k,\,l)|s^0=0,\,\mathcal{M}^1)\right]$, where the factor of $\frac{1}{2}$ is due to Bob sending boxes $0$ and $1$ with equal probability. Suppose that Alice wishes to reveal $b=1$ ($s^c=3$), then $r^c=b$ or $r^c=b\oplus1$. Bob will input $s^{\bar{c}}=1$ and verify that $r^{\bar{c}}=r^c$. In this case Alice's cheating probability equals $\frac{1}{2}\sum_{k,\,l=0,\,1}\left[P(r^1=l,\,(k,\,l)|s^1=1,\,\mathcal{M}^0)+P(r^0=l,\,(k,\,l)|s^0=1,\,\mathcal{M}^1)\right]$. Alice's overall cheating probability is therefore given by
\begin{eqnarray}
& &\frac{1}{4}\sum_{k,\,l=0,\,1}\Bigl[P(r^1=k,\,(k,\,l)|s^1=0,\,\mathcal{M}^0)+P(r^0=k,\, (k, \, l)|s^0=0,\,\mathcal{M}^1)\Bigr.\nonumber\\
& & \Bigl.+P(r^1=l,\,(k,\,l)|s^1=1,\,\mathcal{M}^0)+P(r^0=l,(\,k,\,l)|s^0=1,\,\mathcal{M}^1)\Bigr]\,.
\end{eqnarray}

To obtain Alice's control, we must maximize the above expression under the constraint that the CHSH expectation value is no less than $I_\mathrm{th}$. This translates to the following optimization problem
\begin{eqnarray}
 P_\mathrm{cont}  & = & \frac{1}{4}\max_{\mathcal{Q}}\mathrm{Tr}\biggl(\rho\sum_{c,\, k,\, l=0,\,1}M_{kl}^{c}\bigl(\Pi_{k|0}^{\bar{c}}+\Pi_{l|1}^{\bar{c}}\bigr)\biggr)\nonumber\\
 & \mathrm{s.t.}\quad & \mathrm{Tr}\biggl(\rho\sum_{a,\, b,\, x,\, y=0,\,1}\left(-1\right)^{a\oplus b\oplus xy}\Pi_{a| x}^{0}\Pi_{b| y}^{1}\biggr)\geq I_{\mathrm{th}\,,}\quad \bigl[\Pi_{i| j}^{c},\,\Pi_{k| l}^{\bar{c}}\bigr]=\bigl[M_{ij}^{c},\,\Pi_{k\mid l}^{\bar{c}}\bigr]=\bigl[M_{ij}^{c},\, M_{kl}^{\bar{c}}\bigr]=0\,,\nonumber\\
 &  & \Pi_{i\mid j}^{c}\succeq0\,,\quad M_{ij}^{c}\succeq0\,,\quad\sum_{i=0,\,1}\Pi_{i|j}^{c}=
 \mathds{1}\,,\quad\sum_{i,\, j=0,\,1}M_{ij}^{c}=\mathds{1}\,,
 \label{maximization}\end{eqnarray}
where $\mathcal{Q}=\bigl\{\mathcal{H}^c,\,\rho,\, \{\Pi_{i|j}^c\},\,\mathcal{M}^c\bigr\}_c$ and $\Pi_{r|s}^{c}$ is the POVM element corresponding to inputting $s$ into box $c$ and obtaining the output $r$. Problems of this type can be relaxed to a hierarchy of semi-definite programming (SDP) problems, using the method introduced in \cite{Navascues07,Pironio10B}. This hierarchy provides increasingly tighter upper bounds on the solution of the original problem, which are guaranteed to converge to it at a sufficiently high order. We have solved the second order SDP relaxation of Eq. (\ref{maximization}). In the next subsection we present a cheating strategy which saturates it (up to $10^{-8}$ -- the numerical accuracy of the SDP solver), implying that the second order relaxation already converges. Fig. 1 presents Alice's control as a function of the CHSH expectation value.

\begin{figure}[t!]
\centering
\includegraphics[scale=0.40]{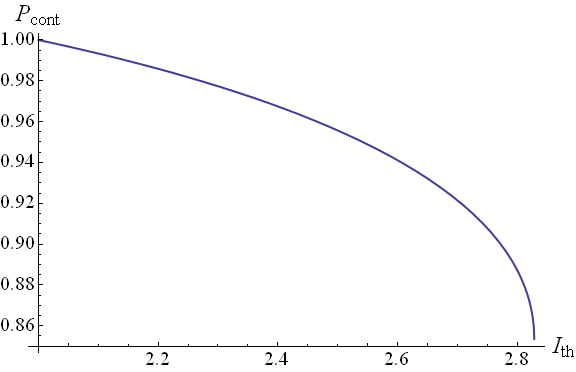}
\caption{Alice's control as a function of $I_\mathrm{th}$ in the asymptotic limit. The curve was obtained from Eqs. (\ref{Optimal Strategy Pcont})-(\ref{Optimal Strategy phi}). The curve saturates the second order relaxation of Eq. (\ref{maximization}) up to $10^{-8}$ -- the numerical accuracy of the SDP solver.}
\label{DishonestAliceGraph}
\end{figure}

\subsection{Alice's optimal cheating strategy in the asymptotic limit}

We present below an optimal cheating strategy, in which it suffices for Alice to perform a single two-outcome measurement, rather than a four-outcome one as described in the previous subsection.
The strategy proceeds as follows. Alice prepares the boxes such that each contains one qubit out of a pair in the maximally
entangled state $\left|\phi^+\right\rangle =\frac{1}{\sqrt{2}}\left(\left|00\right\rangle +\left|11\right\rangle \right)$. Box $0$ is prepared such that inputting $0$ and $1$ gives rise to the measurements $\sigma_{2\theta}$ and
$\sigma_z$, respectively, where $\sigma_\alpha=\cos\alpha\sigma_z+\sin\alpha\sigma_x$. Box $1$ is prepared such that inputting $0$ and $1$ gives rise to the measurements $\sigma_{2\theta-\varphi}$ and $\sigma_{4\theta-\varphi}$, respectively.
If Alice receives box $0$ ($1$) she measures $\sigma_{3\theta - \varphi}$ ($\sigma_\theta$). That is, she always measures along an axis midway between Bob's measurement axes in $zx$-plane (see Fig. 2). She then sends Bob values of $b$ and $r^c$ equal to the result of her measurement. Pairs of measurements along axes, differing by an angle of $\theta$, in the $zx$-plane (since $|\phi^+\rangle$ is invariant under rotations in the $zx$-plane) give rise to correlated outcomes with probability $\cos^2\bigl(\frac{\theta}{2}\bigr)$.  Therefore, irrespectively of whether Alice reveals $0$ or $1$ (or, what is the same thing, whether Bob inputs $0$ or $1$), her cheating probability equals
\begin{equation}
P_\mathrm{cont}=\cos^2\Bigl(\frac{\theta}{2}\Bigr)\,.
\label{Optimal Strategy Pcont}
\end{equation}

Of course the values of $\theta$ and $\varphi$ are restricted by the constraint on the value of the CHSH violation. For the measurements above we have
\begin{equation}
I=\langle \phi^+ | \sigma_{2 \theta} \otimes \sigma_{2 \theta - \varphi}  + \sigma_{2 \theta} \otimes \sigma_{4 \theta - \varphi} + \sigma_z \otimes \sigma_{2 \theta - \varphi}  - \sigma_z \otimes \sigma_{4 \theta - \varphi} | \phi^+ \rangle =2\cos\left(2\theta-\varphi\right)-\cos\left(4\theta-\varphi\right)+\cos\left(\varphi\right)\,.
\label{Optimal Strategy CHSH}
\end{equation}
For a given value of $\theta$ the maximum violation is obtained for
\begin{equation}
\varphi_\mathrm{opt}=\arccos\biggl(2\frac{\cos\left(2\theta\right)+\sin^{2}\left(2\theta\right)}{\sqrt{6-2\cos\left(4\theta\right)}}\biggr)\,.
\label{Optimal Strategy phi}
\end{equation}
By plugging $\varphi_\mathrm{opt}$ into Eq. (\ref{Optimal Strategy CHSH}), and using Eq. (\ref{Optimal Strategy Pcont}) to obtain $\theta$ as a function of $P_\mathrm{cont}$, we obtain $I$ as a function of $P_\mathrm{cont}$. The resulting curve saturates the SDP obtained curve in Fig. 1.

\begin{figure}[t!]
\centering
\includegraphics[scale=0.5]{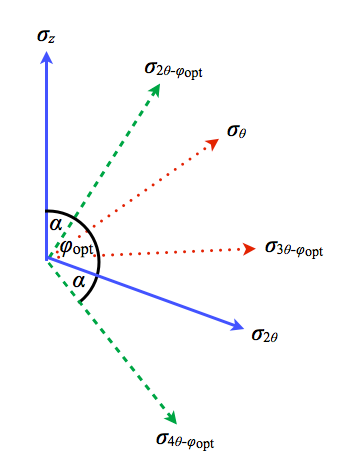}
\caption{Schematic representation of the alignment of the measurement axes in Alice's optimal cheating strategy. The solid (dashed) axes correspond to Bob's measurement on box $0$  ($1$). The dotted axes correspond to Alice's measurements (Alice always measures midway between Bob's axes). The axes all lie in the $zx$-plane. $\alpha = 2\theta -\varphi_{\mathrm{opt}}$. $\varphi_{\mathrm{opt}}$ and $\theta$ are related via Eq. (\ref{Optimal Strategy phi})}.
\label{OptimalStrategyA}
\end{figure}
\subsection{Alice's control in the general case of an arbitrary number of tests}
For any given value of $n$, Alice's control is a function of the CHSH expectation value $E(I(W_{n+1})|\mathrm{\mathbf{w}}_n)$ characterizing the behavior of the devices at step $n+1$ given the history $\mathrm{\mathbf{w}}_n$. Alice's control can therefore be expressed as
\begin{eqnarray}
P_{\mathrm{cont}} & = & \frac{1}{N}\sum_{n=1}^{N-1}\sum_{\{\mathrm{\mathbf{w}}_n\} }P(\mathrm{\mathbf{w}}_n)\Theta(\bar{I}_n(\mathrm{\mathbf{w}}_n)-I_\mathrm{th})C(E(I(W_{n+1})|\mathrm{\mathbf{w}}_n))+\frac{1}{N}\sum_{\{\mathrm{\mathbf{w}}_N\} }P(\mathrm{\mathbf{w}}_N)\Theta(\bar{I}_N(\mathrm{\mathbf{w}}_N)-I_\mathrm{th})\nonumber\\
& \leq & \frac{1}{N}\sum_{n=1}^{N-1}\sum_{\{\mathrm{\mathbf{w}}_n\} }P(\mathrm{\mathbf{w}}_n)\Theta(\bar{I}_n(\mathrm{\mathbf{w}}_n)-I_\mathrm{th})C(E(I(W_{n+1})|\mathrm{\mathbf{w}}_n))+\frac{1}{N}\,,
\end{eqnarray}
where $\Theta(\bar{I}_n(\mathrm{\mathbf{w}}_n)-I_\mathrm{th})$ is the unit
step function, ensuring that only histories, such that the observed CHSH violation is no less than the threshold $I_\mathrm{th}$, contribute. The partitioning of the sum into two is due to the fact that the boxes may have internal counters keeping track of the number of times they have been tested. Since the $N+1\,$th use of the boxes, if occurring at all (i.e. if Bob picks $N$), necessarily occurs in the reveal phase, it is never part of the CHSH testing. Therefore, in an optimal cheating strategy Alice will program the boxes such that in their $N+1\,$th use they behave deterministically.

For each history $\mathrm{\mathbf{w}}_n$ with $n\leq N-2$
we can define the set of all \emph{compatible} histories $\mathrm{\mathbf{w}}_{N-1}$ that could have occurred had Bob carried out $N-1$ repetitions instead of $n$. Alice's control can therefore be re-expressed as
\begin{equation}
P_{\mathrm{cont}} \leq \sum_{\{ \mathrm{\mathbf{w}}_{N-1}\} }P(\mathrm{\mathbf{w}}_{N-1})\frac{1}{N}\sum_{n=1}^{N-1}\Theta(\bar{I}_n(\mathrm{\mathbf{w}}_n)-I_\mathrm{th})C(E(I(W_{n+1})|\mathrm{\mathbf{w}}_n))+\frac{1}{N}\,.
\end{equation}

Let $K(\mathrm{\mathbf{w}}_{N-1})$ denote the last repetition, up to and including the $N-1\,$th repetition, of the compatible history $\mathrm{\mathbf{w}}_{N-1}$
for which the observed CHSH violation is no less than $I_\mathrm{th}$, i.e.
\begin{equation}
K(\mathrm{\mathbf{w}}_{N-1})=\max_{k\leq N-1}\{k |  \bar{I}_k(\mathrm{\mathbf{w}}_k)\geq I_\mathrm{th}\}\,.
\end{equation}
We can bound Alice's control probability as follows:
\begin{eqnarray}
P_\mathrm{cont} & \leq & \sum_{\{\mathrm{\mathbf{w}}_{N-1}\}}P(\mathrm{\mathbf{w}}_{N-1})\frac{1}{N}\sum_{n=1}^{N-1}\Theta(\bar{I}_n(\mathrm{\mathbf{w}}_{n})-I_\mathrm{th})C(E(I(W_{n+1})|\mathrm{\mathbf{w}}_{n}))+\frac{1}{N}\nonumber \\
 & = & \sum_{k=1}^{N-1}\sum_{\{\mathrm{\mathbf{w}}_{N-1}|K(\mathrm{\mathbf{w}}_{N-1})=k\}}P(\mathrm{\mathbf{w}}_{N-1})\frac{1}{N}\sum_{n=1}^{N-1}\Theta(\bar{I}_n(\mathrm{\mathbf{w}}_{n})-I_\mathrm{th})
 C(E(I(W_{n+1})|\mathrm{\mathbf{w}}_n))+\frac{1}{N}\nonumber \\
 & \leq & \sum_{k=1}^{N-1}\sum_{\{\mathrm{\mathbf{w}}_{N-1}|K(\mathrm{\mathbf{w}}_{N-1})=k\}}P(\mathrm{\mathbf{w}}_{N-1})\frac{1}{N}\sum_{n=1}^{k}C(E(I(W_{n+1})|\mathrm{\mathbf{w}}_{n}))+\frac{1}{N}\,,\end{eqnarray}
where we have used the fact that $\Theta(\bar{I}_n(\mathrm{\mathbf{w}}_n)-I_\mathrm{th})=0$ for all $n$ such that $K(\mathrm{\mathbf{w}}_{N-1}) < n \leq N-1 $; the inequality being due to the possibility of histories for which $\Theta(\bar{I}_n(\mathrm{\mathbf{w}}_n)-I_\mathrm{th})=0$ for at least one value of $n < K(\mathrm{\mathbf{w}}_{N-1})$.

Defining $K_{0}=\lceil (N-1) C(I_\mathrm{th}) \rceil $, we have
\begin{eqnarray}
P_\mathrm{cont} &  \leq  & \sum_{k=1}^{K_{0}-1}\sum_{\{\mathrm{\mathbf{w}}_{N-1}|K(\mathrm{\mathbf{w}}_{N-1})=k\}}P(\mathrm{\mathbf{w}}_{N-1})\frac{k}{N} +\sum_{k=K_{0}}^{N-1} \sum_{\{\mathrm{\mathbf{w}}_{N-1}|K(\mathrm{\mathbf{w}}_{N-1})=k\}}P(\mathrm{\mathbf{w}}_{N-1})\frac{k}{N}C\Bigl(\frac{1}{k}\sum_{n=1}^k E(I(W_{n+1})|\mathrm{\mathbf{w}}_n)\Bigr) +\frac{1}{N}\nonumber\\
 &  \leq &   \sum_{k=1}^{K_{0}-1}\sum_{\{\mathrm{\mathbf{w}}_{N-1}|K(\mathrm{\mathbf{w}}_{N-1})=k\}}P(\mathrm{\mathbf{w}}_{N-1})\frac{(N-1)C(I_\mathrm{th})}{N}\nonumber\\
 & & +\sum_{k=K_{0}}^{N-1} \sum_{\{\mathrm{\mathbf{w}}_{N-1} |K(\mathrm{\mathbf{w}}_{N-1})=k,\,\mathrm{\mathbf{w}}_{N-1}\notin\pi_{k}(\varepsilon)\}}P(\mathrm{\mathbf{w}}_{N-1})\frac{N-1}{N}C\Bigl(\frac{1}{k}\sum_{n=1}^k E(I(W_{n+1})|\mathrm{\mathbf{w}}_n)\Bigr)
\nonumber\\
& &
+\sum_{k=K_{0}}^{N-1}\sum_{\{\mathrm{\mathbf{w}}_{N-1} |K(\mathrm{\mathbf{w}}_{N-1})=k,\,\mathrm{\mathbf{w}}_{N-1}\in\pi_{k}(\varepsilon)\}}P(\mathrm{\mathbf{w}}_{N-1})\frac{N-1}{N}C\Bigl(\frac{1}{k}\sum_{n=1}^k E(I(W_{n+1})|\mathrm{\mathbf{w}}_n)\Bigr)+\frac{1}{N}\qquad \forall \,\varepsilon \geq 0\,,
\end{eqnarray}
where in the second sum in the first line we have used the concavity of $C$, in the first sum in the second line the fact that $K_{0} - 1\leq (N-1)C(I_\mathrm{th})$, and where $\pi_{k}\left(\varepsilon\right)$ is
 the set of all histories satisfying 
\begin{equation}
\bar{I}_k\left(\mathbf{w}_{k}\right)-\frac{1}{k}\sum_{n=1}^{k}E\left(I\left(W_{n}\right)|\mathbf{w}_{n-1}\right)\geq\varepsilon\,.
\end{equation}
In Appendix D we show that the probability of occurrence of $\pi_k(\varepsilon)$ is bounded by \begin{equation}
P\left(\pi_{k}\left(\varepsilon\right)\right)\leq\exp\biggl(-\frac{k\varepsilon^{2}}{2D^{2}}\biggr)\,,\end{equation}
where $D=4+2\sqrt{2}$, and so  \begin{equation}
\sum_{k=K_{0}}^{N-1}P\left(\pi_{k}\left(\varepsilon\right)\right)\leq\frac{\exp\bigl(-\frac{K_{0}\varepsilon^{2}}{2D^{2}}\bigr)-\exp\bigl(-\frac{N\varepsilon^{2}}{2D^{2}}\bigr)}{1-\exp\left(-\frac{\varepsilon^{2}}{2D^{2}}\right)}=Q(\varepsilon)\,.\end{equation}
\begin{figure}[t!]
\centering
\includegraphics[scale=0.4]{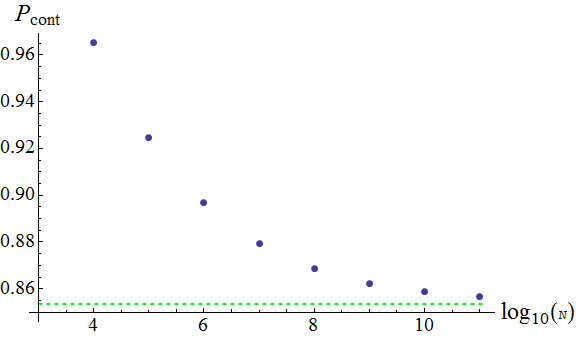}
\caption{Upper bound on Alice's control as a function of $\log_{10}N$. The curve presents the results of numerical solutions of Eq. (\ref{Bound Pcont}) for different values of $N$, given $I_\mathrm{th}=2\sqrt{2} (1-\frac{1}{\sqrt{N}})$. In the limit $N\rightarrow\infty$ Alice's control tends to the asymptote $\cos^2 \bigl(\frac{\pi}{8}\bigr)\simeq 0.854$ (represented by the dashed curve).}
\end{figure}
Making use of this last inequality, we finally get that
\begin{eqnarray}
P_\mathrm{cont} &  \leq   & \frac{N-1}{N}\sum_{k=1}^{K_{0}-1}\sum_{\{\mathrm{\mathbf{w}}_{N-1}|K(\mathrm{\mathbf{w}}_{N-1})=k\}}P(\mathrm{\mathbf{w}}_{N-1})C(I_\mathrm{th})\nonumber\\
& &+\frac{N-1}{N}\min_{\varepsilon \geq 0}\biggl[\sum_{k=K_{0}}^{N-1}\sum_{\{\mathrm{\mathbf{w}}_{N-1} |K(\mathrm{\mathbf{w}}_{N-1})=k,\,\mathrm{\mathbf{w}}_{N-1}\notin\pi_{k}(\varepsilon)\}}P(\mathrm{\mathbf{w}}_{N-1})C(\bar{I}_k(\mathrm{\mathbf{w}}_{k})-\varepsilon)\nonumber\\
& & \biggl.+\sum_{k=K_{0}}^{N-1}\sum_{\{\mathrm{\mathbf{w}}_{N-1} |K(\mathrm{\mathbf{w}}_{N-1})=k,\,\mathrm{\mathbf{w}}_{N-1}\in\pi_k(\varepsilon)\}}P(\mathrm{\mathbf{w}}_{N-1})\biggr]+\frac{1}{N}\nonumber \\
 &  \leq   & \frac{N-1}{N}\min_{\varepsilon \geq 0}\biggl[\sum_{k=1}^{K_{0}-1}\sum_{\{\mathrm{\mathbf{w}}_{N-1}|K(\mathrm{\mathbf{w}}_{N-1})=k\}}P(\mathrm{\mathbf{w}}_{N-1})C(I_\mathrm{th}-\varepsilon) \biggr. \nonumber\\
 & & \biggl.+\sum_{k=K_{0}}^{N-1}\sum_{\{\mathrm{\mathbf{w}}_{N-1} |K(\mathrm{\mathbf{w}}_{N-1})=k\}}P(\mathrm{\mathbf{w}}_{N-1})C(I_\mathrm{th}-\varepsilon)\nonumber\\
 & &+\sum_{k=K_{0}}^{N-1}\sum_{\{\mathrm{\mathbf{w}}_{N-1} |K(\mathrm{\mathbf{w}}_{N-1})=k,\,\mathrm{\mathbf{w}}_{N-1}\in\pi_k(\varepsilon)\}}P(\mathrm{\mathbf{w}}_{N-1})(1-C(I_\mathrm{th}-\varepsilon))\biggr]+\frac{1}{N}\nonumber \\
 & \leq & \frac{N-1}{N}\min_{\varepsilon \geq 0} \Bigl[C(I_\mathrm{th} -\varepsilon)+(1-C(I_\mathrm{th}-\varepsilon))Q(\varepsilon)\Bigr]+\frac{1}{N}\,.
\label{Bound Pcont} 
\end{eqnarray}
Note that if we choose the behavior of $\varepsilon$ such that in the limit $N\rightarrow \infty$ it decays more slowly than $N^{-1/2}$, then $\lim_{N\rightarrow \infty}Q(\varepsilon)\rightarrow 0$ and the bound tends to $C(I_\mathrm{th})$. For finite $N$ it seems unlikely that the bound is saturable, since $Q(\varepsilon)$ is non-vanishing. Fig. 3 presents the results of numerical solutions of Eq. (\ref{Bound Pcont}) for different values of $N$ for $I_\mathrm{th}=2 \sqrt{2} -\frac{1}{\sqrt{N}}$. In particular, in the limit $N\rightarrow \infty$ Alice's control tends to $\cos^2\left(\frac{\pi}{8}\right)$, recovering the result of \cite{Silman11}.

\section{Summary}

Distrustful cryptography presents unique challenges in device-independent settings, which are absent in non-distrustful cryptographic tasks, such as quantum key-distribution. In particular, since the parties do not trust each other and may have conflicting goals, they cannot work together to certify the presence of nonlocality. In \cite{Silman11,Aharon14} this problem was circumvented by making use of the pseudo-telepathic nature of GHZ correlations, but pseudo-telepathy is absent in a CHSH setting. In this work we have shown that pseudo-telepathy is not essential for doing device-independent distrustful cryptography. This was achieved by reformulating the device-independent bit commitment protocol of \cite{Silman11}, such that it relies on sequential testing of the CHSH inequality (instead of the single-shot testing of GHZ correlations), but (in the asymptotic limit) nevertheless achieves the same security. The security analysis was therefore carried out in the most general settings, where the devices may have long-term quantum memory.

Strictly speaking, the protocol we have presented is not a bit commitment protocol since Alice cannot choose the reveal time at will. This by itself is not necessarily a problem. For example, it does not prevent the protocol from being used to implement coin flipping. However, if we would like Alice to have the freedom to choose the reveal time, then we can do so either, as shown in Appendix B, at the price of increasing her control, or, as shown in Appendix C, at the price of using additional resources, i.e. by working in the large office scenario, where the parties have access to many pairs of boxes, which can be measured in parallel.

Our work opens the door for real-life implementation of device-independent bit commitment and coin flipping. The protocol of \cite{Silman11} requires the ability to reliably produce particles in a GHZ state and to store, manipulate, and transmit them while maintaining their coherence. The protocol presented here, on the other hand, only requires manipulation of bipartite entanglement which is simpler given state-of-the-art technology.

Finally, we point out that the techniques developed in this work are not especially tailored for device-independent bit commitment, and we expect them to be useful, and possibly even essential, for other distrustful cryptographic tasks, such as non-bit commitment-based device-independent coin flipping, and device-independent oblivious transfer.

\begin{acknowledgments}
S.P., S.M., and J.S. acknowledge financial support from the European Union under the
projects QCS, QALGO, and DIQIP, and from the F.R.S.-FNRS under the
project DIQIP. S.P. acknowledges support from the Brussels-Capital Region through a BB2B grant. S.P. is a Research Associate of the Fonds de la Recherche Scientifique F.R.S.-
FNRS (Belgium). J.S. was a postdoctoral researcher of the Fonds de la Recherche Scientifique F.R.S.- FNRS (Belgium) at the time this research was carried out. N.A acknowledges support from the BSF (grant no. 32/08) and the
Niedersachsen-Israeli Research Cooperation Program.
The Matlab toolboxes YALMIP  \cite{YALMIP} and SeDuMi \cite{SeDuMi} were
used to solve the SDP problem Eq. (\ref{maximization}).
\end{acknowledgments}

\section*{Appendix A}
We present a device-independent bit commitment protocol using a PR box \cite{Popescu94}.\\

We have seen that reformulating the GHZ-based protocol of \cite{Silman11} to be CHSH-based comes at the price of pseudo-telepathy. Indeed, quantum theory does not allow for pseudo-telepathy in a two-party, two-input setting \cite{Gisin07}. However, in a post-quantum world -- in which both \emph{dishonest and honest} parties are restricted only by the no-signaling constraints -- pseudo-telepathy is restored. It is interesting to ask what would happen to our protocol if we were to adapt it to such a world. On the one hand, we might expect such a world to offer less security since a dishonest party would now have access to stronger correlations. On the other hand, we might expect the converse, since the protocol itself can be modified to make use of these stronger correlations. We will see that on the balance this allows for more security\footnote{The authors of \cite{Buhrman06} presented a perfect bit commitment protocol, assuming that honest parties have access to PR boxes and dishonest parties cannot tamper with them. We do not make these assumptions.}.

A PR box is a post-quantum, bipartite, two-input, two-output resource, which achieves the algebraic bound of the
CHSH inequality, while at the same time satisfying the no-signaling constraints. Up to local relabeling of the inputs and outputs, the PR box  satisfies
\begin{equation}
r^0\oplus r^1=s^0\cdot s^1\,,\qquad r^i,\,s^i\in\{0,\,1\}\,.
\label{PRBox}
\end{equation}
In the following, it will be convenient to think of the PR box as consisting of a pair of two-input, two-output boxes, one in the possession of Alice, and the other in the possession of Bob.

The PR-based protocol is essentially a simplified version of our earlier protocols with the first step (statistical estimation of the CHSH violation and random selection of the box used to encode Alice's commitment) omitted, and with the verification of nonlocal correlations and Alice's commitment being performed at the same time. This last possibility follows from pseudo-telepathy, as in the protocol of \cite{Silman11}.

We assume that at the start of the protocol Alice has box $0$ and Bob has box $1$.
The protocol proceeds as follows:

\begin{enumerate}
\item \emph{Commit phase} - Alice inputs into her box the
value of the bit she wishes to commit. She then selects uniformly at random
a classical bit $a$, and sends Bob another classical bit,
$q=r^0\oplus (a s^0)$, as a token of her commitment.
\item \emph{Reveal phase} - Alice sends Bob $s^0$ and $r^0$. Bob checks whether
$q=r^0$ or $q=r^0\oplus s^0$. If both relations are not satisfied, he aborts.
Otherwise, he picks an input $s^1$ uniformly at random and verifies that $r^0\oplus r^1=s^0\cdot s^1$.
If this last test fails, he aborts.
\end{enumerate}

\subsubsection*{Alice's security}

We recall that in the quantum case (both in the GHZ-based and CHSH-based formulations)
Alice's security relies only on the no-signaling constraints. Since we are still working in a non-signaling setting, Alice's security will remain unchanged, i.e. Bob's information gain is upper-bounded by $\frac{3}{4}$. The proof proceeds exactly as in Subsection IV.A, except that the instead of inputting $2$ and $3$, Alice inputs $0$ and $1$.

One optimal strategy for Bob is to assume that $q=r^0$, and
input $s^1=1$, obtaining an output $r^1$. He then guesses $g=r^0\oplus r^1$.
When $s^0=0$, $q=r^0$ and Bob's guess, $g=r^0\oplus r^1=s^0$,
is correct. When $s^0=1$ and $a=0$, $q=r^0$ and Bob's guess, $g=r^0\oplus r^1=s^0$, is again
correct. However, when $s^0=1$ and $a=1$, $q=r^0\oplus 1$ and  Bob's guess, $g=r^0\oplus r^1\oplus1=s^0\oplus1$,
is wrong. Since Alice is honest she picks $a$ uniformly at random, implying that $P_\mathrm{gain}=\frac{3}{4}$.

\subsubsection*{Bob's security}

Recall that in a device-independent scenario dishonest Alice can prepare
the boxes in any state she wishes, possibly entangled
with ancillary systems in her possession. Since in the commit phase Alice sends
a classical bit $q$ as a token of her commitment, without receiving
any information from Bob, without loss of generality we may assume
that she decides on the value of $q$ before the start of the protocol, and accordingly
prepares the boxes to maximize $P_\mathrm{cont}$. Furthermore, since Alice's
cheating probability is invariant under the simultaneous relabeling $q\rightarrow q\oplus 1$ and
$r^0\rightarrow r^0\oplus1$, no value of $q$ is preferable,
and we may assume that she sends $q=0$.

Suppose now that Alice wishes
to reveal $0$. Since $s^0=0$, it follows that Alice must send $r^0=0$ as Bob will first check whether $r^0=0$ or not. Bob will then test whether the PR box correlations, Eq. (\ref{PRBox}), are satisfied:
 Bob will uniformly at random pick a value of
$s^1$ and verify that $r^1=0$.
Alice's control in this case equals $\frac{1}{2}\left[ P \left( r^1 = 0 | s^1 = 0 \right) + P \left( r^1 = 0 | s^1 = 1 \right) \right]$.
Suppose now that Alice wishes to reveal $1$, then
both values of $r^{1}$ are possible, and the only relevant
test is whether the PR box correlations are satisfied. Alice's control
in this case equals $\frac{1}{2}\sum_{r^0=0,\,1} \left[ P \left( r^0,\,r^1=r^0 | s^0 = 1,\,s^1 = 0 \right) + P \left( r^0,\,r^1 = r^0 \oplus 1 | s^0 = 1,\, s^1 = 1 \right) \right]$.
Alice's overall control is obtained by maximizing

\begin{eqnarray}
 &  & \frac{1}{4}\left[P(r^{1}=0|s^{1}=0)+P(r^{1}=0|s^{1}=1)\right.\nonumber\\
 & + & \sum_{r^{0}=0,\,1}\bigl(P(r^{0},\, r^{1}=r^{0}\mid s^{0}=1,\, s^{1}=0)+\left.P(r^{0},\, r^{1}=r^{0}\oplus1\mid s^{0}=1,\, s^{1}=1)\bigl)\right]\nonumber\\
 & = & \frac{1}{4}\left[P(r^{1}=0|s^{1}=0)+P(r^{1}=0|s^{1}=1)\right.\nonumber\\
 & + & P(r^{0}=0,\, r^{1}=0\mid s^{0}=1,\, s^{1}=0)+P(r^{0}=0,\, r^{1}=1\mid s^{0}=1,\, s^{1}=1)\nonumber\\
 & + & P(r^{0}=1,\, r^{1}=1\mid s^{0}=1,\, s^{1}=0)+\left.P(r^{0}=1,\, r^{1}=0\mid s^{0}=1,\, s^{1}=1)\right]
\end{eqnarray}
and is easily seen to be no greater than $\frac{3}{4}$.

Alice's optimal strategy is to prepare Bob's box such that $P(r^1=0|s^1=0)=P(r^1=0|s^1=1)=1$, i.e. a classical box. 

\section*{Appendix B}

In this appendix we consider a modification of the protocol, such that the reveal time can be chosen at will. This comes at the price of increasing Alice's control. The protocol proceeds as follows:
\begin{enumerate}
\item \emph{Random selection} -- At time $t^a <t_1$ Bob picks uniformly at random, and in private, a number $n\in\{1,\,\dots,\,N\}$  and two input strings $\mathrm{\mathbf{s}}^0_n\in\{0,\,1\}^n$ and $\mathrm{\mathbf{s}}^1_n\in\{0,\,1\}^n$. At each of the $n$ times $t_i$ he feeds $s^0_i$ and $s^1_i$ into boxes $0$ and $1$, respectively.
He uses the corresponding output strings $\mathrm{\mathbf{r}}^0_n$ and $\mathrm{\mathbf{r}}^1_n$ to compute the observed CHSH violation, $\bar{I}_n(\mathrm{\mathbf{w}}_n)$, and compares it to some previously agreed threshold $I_\mathrm{th}$.
If $\bar{I}_n(\mathrm{\mathbf{w}}_n) < I_\mathrm{th}$, he aborts the protocol.
Otherwise, he flips two classical coins. Denote their outcomes by $c$ and $d$. At time $t^b  < t_{n+1}$ he
sends box $c$ to Alice. At time $t_{n+1}$, he inputs $s_{n+1}^{\bar{c}}=d$ into box $\bar{c}$.
\item \emph{Commit phase} -- Let $b$ be the value of the bit Alice wishes to commit. Alice inputs $s^c_{n+1}=b+2$ into her
box. She then selects uniformly at random a classical bit $a$, and at time $t^c>t^b$
sends Bob the classical bit $q=r^c_{n+1}\oplus a b$ as a token
of her commitment.
\item \emph{Reveal phase} -- At any time of her choosing $t>t^c$ Alice sends Bob $b$
and $r^c_{n+1}$. Bob checks whether $q=r^c_{n+1}$ or $q=r^c_{n+1}\oplus b$. If both relations are not satisfied, he aborts. Otherwise, if  $s^{\bar{c}}_{n+1}=s^c_{n+1}-2$ (i.e. $d=b-2$) he verifies that $r^{\bar{c}}_{n+1}=r^c_{n+1}$.
If this last test fails, he aborts.
\end{enumerate}

\subsubsection*{Alice's security}

Clearly, the analysis of Alice's security remains the same as in the original protocol.

\subsubsection*{Bob's security}

Bob's random choice of the $n+1\,$th input into box $\bar{c}$ implies that half the time he will not be able to carry out the last test in the reveal phase and will accept whatever value dishonest Alice reveals, independently of her cheating strategy. The remaining half of the time, Bob's actions will precisely be identical to those analyzed previously (in particular, the last input is introduced into box $\bar{c}$ at time $t_{n+1}$, and thus, unless $n=N$, this box cannot tell whether it is measured in the random selection phase or the reveal phase). Alice's control is therefore modified as follows: $P_\mathrm{cont}\rightarrow \frac{1}{2}(P_\mathrm{cont}+1)$. 

\section*{Appendix C}

In this appendix we consider the large office scenario, where Bob has a large number of pairs of boxes, and where the boxes in each pair can not only be prevented from communicating with one another, but also with the other pairs. While impractical, this scenario allows us to modify the protocol such that the reveal time can be chosen at will while maintaining the same level of security.\\  

Before we begin, we adapt the notation to this new scenario. We consider $N+1$ pairs\footnote{In the original protocol the maximum number of uses of each box equals $N+1$ ($n\in\{1,\,N\}$ for the CHSH estimation and $1$ for the commitment).} of four-input, $\{0,\,1,\,2,\,3\}$, two-output, $\{0,\,1\}$, boxes. The random variables designating the input and
output of the $i\,$th box ($i\in\{0,\,1\}$) of the $k\,$th pair will be labeled by $S^i_k$ and $R^i_k$,
respectively, with a specific realization  being
labeled by lower-case letters $s^i_k$ and $r^i_k$. We also define $W_k=\{S^0_k,\,S^1_k,\,R^0_k,\,R^1_k\}$ and $\mathbf{W}_n=\{W_1,\,\dots,\,W_n\}$. Finally, we define the random strings $\mathbf{S}^i_{\bar{n}}=\{S^i_1,\,\dots,\,S^i_{n-1},\,S^i_{n+1},\,\dots,\,S^i_{N+1}\}$, $\mathbf{R}^i_{\bar{n}}=\{R^i_1,\,\dots,\,R^i_{n-1},\,R^i_{n+1},\,\dots,\,R^i_{N+1}\}$, and $\mathbf{W}_{\bar{n}}=\{W_1,\,\dots,\,W_{n-1},\,W_{n+1},\,\dots,\,W_{N+1}\}$.\\

The protocol proceeds as follows:

\begin{enumerate}
\item \emph{Random selection} -- Bob picks uniformly at random and \emph{in private} a number $n\in\{1,\,\dots,\,N+1\}$ and a classical bit $c$. He sends box $c$ of the $n\,$th pair to Alice.
\item \emph{Commit phase} -- Let $b$ be the value of the bit Alice wishes to commit. Alice inputs $s^c_n=b+2$ into her
box. She then selects uniformly at random a classical bit $a$, and sends Bob the classical bit $q=r^c_{n+1}\oplus a b$ as a token
of her commitment.
\item \emph{Reveal phase} -- Alice sends Bob $b$
and $r^c_{n}$.  Bob checks whether $q=r^c_{n}$ or $q=r^c_{n}\oplus b$. If both relations are
not satisfied, he aborts. Otherwise, he picks uniformly at random two input strings
 $\mathrm{\mathbf{s}}^0_{\bar{n}}\in\{0,\,1\}^N$
and $\mathrm{\mathbf{s}}^1_{\bar{n}}\in\{0,\,1\}^N$, which he feeds into the corresponding boxes. 
\emph{At the same time} he feeds $s^{\bar{c}}_{n}=s^c_{n}-2=b$ into box $\bar{c}$ of the $n\,$th pair. If $r^{\bar{c}}_{n+1} \neq r^c_{n+1}$, he aborts.  
Else, he uses the corresponding output strings $\mathrm{\mathbf{r}}^0_{\bar{n}}$ and $\mathrm{\mathbf{r}}^1_{\bar{n}}$ to compute the observed CHSH violation $\bar{I}_{\bar{n}}(\mathbf{w}_{\bar{n}})=\frac{1}{N}\sum_{k\neq n}I(w_k)$ and compares it to some previously agreed threshold $I_\mathrm{th}$.
If this last test fails, he aborts.
\end{enumerate}

\subsubsection*{Alice's security}
Clearly, the analysis of Alice's security remains the same as in the original protocol.

\subsubsection*{Bob's security}

We will not derive here the dependence of Alice's control on the number of pairs $N+1$. Instead, we will show that it is upper-bounded by that of the original protocol (Section III). To see this, consider another protocol, identical to the one above in all except that instead of using the inputs and outputs of all the pairs (bar the one chosen for the commitment) to estimate the CHSH violation, Bob uses only those of the first $n-1$ pairs (Alice is of course aware of this and of the numbering of the pairs). Clearly, the new protocol can only increase Alice's control.

Now we note that this protocol would be identical to that of the sequential case, up to the fact that the reveal time can be chosen at will\footnote{Unlike in the sequential case, timing issues do not arise here since  the measurements for the CHSH estimation and on box $\bar{c}$ of pair $n$ are simultaneous.}, if box $i$ of pair $k$ were to have full information about the inputs and outputs of all the $i\,$th boxes of the first $k-1$ pairs. Clearly, such a modification can only increase Alice's control.

We therefore conclude that Alice's control in the sequential case provides an upper bound on her control in the large office scenario.

\section*{Appendix D}

Let

\begin{equation}
\Delta_k(\mathrm{\mathbf{W}}_{k})=\bar{I}_k(\mathrm{\mathbf{W}}_{k}) - \frac{1}{k}\sum_{n=1}^{k}E(I(W_n)|\mathrm{\mathbf{W}}_{n-1})=\frac{1}{k}\sum_{n=1}^{k}\bigl(I(W_n) - E(I(W_n)|\mathrm{\mathbf{W}}_{n-1})\bigr)\,,\qquad k\leq N-1\,.
\end{equation}
It is straightforward to show that $Z_k(\mathrm{\mathbf{W}}_{k})=k\Delta_k(\mathrm{\mathbf{W}}_{k})$ is a martingale (i.e. $E(Z_{k+1}(\mathrm{\mathbf{W}}_{k+1})|\mathrm{\mathbf{W}}_k))=Z_k(\mathrm{\mathbf{W}}_{k})$). Moreover, for any history $\mathbf{w}_{k}$ we have that \begin{equation}
\bigl|Z_{k+1}\left(\mathbf{w}_{k+1}\right)-Z_{k}\left(\mathbf{w}_{k}\right) \bigr|\leq{D}\,,\end{equation}
where $D=4+2\sqrt{2}=\left(1-\cos^2(\frac{\pi}{8})\right)^{-1}$; the $4$ coming from the $I\left(\mathrm{\mathbf{w}}_{k+1}\right)$ term
and the $2\sqrt{2}$ from the $E\left(I\left(W_{k+1}\mid\mathrm{\mathbf{w}}_{k}\right)\right)$ term.
The Azuma-Hoeffding inequality \cite{Azuma67} then tells us that
\begin{equation}
P(\pi_k(\varepsilon))\leq\exp\Bigl(-\frac{k\varepsilon^{2}}{2D^{2}}\Bigr)\,,\qquad k\leq N-1\,,
\end{equation}
where $\pi_{k}\left(\varepsilon\right)$ is defined to be the union of all 
histories $\mathrm{\mathbf{w}}_{k}$ satisfying $
\Delta_{k}\left(\mathbf{w}_{k}\right)\geq\varepsilon$.

\end{document}